\documentclass[sigconf,screen]{acmart}

\usepackage{soul}
\usepackage{hyperref}
\usepackage{multirow}
\usepackage{listings}
\usepackage{fancybox}
\usepackage{enumitem}
\usepackage{graphicx}
\usepackage{wrapfig}
\usepackage{color}
\usepackage{array}
\usepackage{amsmath}
\usepackage{xspace}
\usepackage{url}
\usepackage{tikz}
\usepackage{caption}
\usepackage{pgfplots}
\usepackage{balance}
\usepackage{subcaption}
\pgfplotsset{compat=1.16}
\usepackage{pgf-pie}
\usetikzlibrary{pgfplots.statistics,calc}
\usepackage{algorithm}
\usepackage{algorithmic}
\usepackage{adjustbox}
\usepackage{makecell}

\usepackage{tikz}

\usepackage{tcolorbox}
\makeatletter
\newcommand{\mybox}[1]{%
	\setbox0=\hbox{#1}%
	\setlength{\@tempdima}{\dimexpr\wd0+13pt}%
	\begin{tcolorbox}[boxrule=0.5pt, colback=white, arc=4pt,
		left=6pt,right=6pt,top=6pt,bottom=6pt,boxsep=0pt]
		#1
	\end{tcolorbox}
}

\definecolor{songcolor}{RGB}{191,191,191}

\AtBeginDocument{%
  }

\setcopyright{acmcopyright}
\copyrightyear{2026}
\acmYear{2026}
\acmDOI{XXXXXXX.XXXXXXX}
\acmPrice{15.00}
\acmISBN{978-1-4503-XXXX-X/18/06}
\acmBooktitle{ACM International Conference on the Foundations of Software Engineering (FSE '26), June 5--9, 2026, Montreal, Canada}
\begin{document}

\title[]{Engineering Pitfalls in AI Coding Tools: An Empirical Study of Bugs in Claude Code, Codex, and Gemini CLI}

\author{Ruixin Zhang}
 \authornote{Both authors contributed equally to this research.}
\affiliation{
  \institution{York University}
  \city{Toronto}
  \country{Canada}
}
\email{jason666@my.yorku.ca}

\author{Wuyang Dai}
\authornotemark[1]
\affiliation{
  \institution{York University}
  \city{Toronto}
  \country{Canada}
}
\email{ddai2002@my.yorku.ca}

\author{Hung Viet Pham}
\affiliation{
  \institution{York University}
  \city{Toronto}
  \country{Canada}
}
\email{hvpham@yorku.ca}

\author{Gias Uddin}
\affiliation{
  \institution{York University}
  \city{Toronto}
  \country{Canada}
}
\email{guddin@yorku.ca}

\author{Jinqiu Yang}
\affiliation{
  \institution{Concordia University}
  \city{Montreal}
  \country{Canada}
}
\email{jinqiu.yang@concordia.ca}

\author{Song Wang}
\affiliation{
  \institution{York University}
  \city{Toronto}
  \country{Canada}
}
\email{wangsong@yorku.ca}

\begin{abstract} 


The rapid integration of Large Language Models (LLMs) into software development workflows has given rise to a new class of AI-assisted coding tools, such as Claude-Code, Codex, and Gemini CLIs. While promising significant productivity gains, the engineering process of building these tools, which sit at the complex intersection of traditional software engineering, AI system design, and human-computer interaction, is fraught with unique and poorly understood challenges. 

This paper presents the first empirical study of engineering pitfalls in building such tools, on a systematic, manual analysis of over 3.8K publicly reported bugs in the open-source repositories of three AI coding tools (i.e., Claude-Code, Codex, and Gemini CLI) on GitHub. Specifically, we employ an open-coding methodology to manually examine the issue description, associated user discussions, and developer responses. Through this process, we categorize each bug in multiple dimensions, including bug type, bug location, root cause, and observed symptoms. This fine-grained annotation enables us to characterize common failure patterns and identify recurring engineering challenges. 


Our results show that more than 67\% of the bugs in these tools are related to functionality. In terms of root causes, 37.3\% 
of the bugs stem from API, integration, or configuration errors. Consequently, the most commonly observed symptoms reported by users are API errors (18.3\%), terminal problems (14\%), and command failures (12.7\%). 
These bugs predominantly affect the tool invocation (37.6\%) and command execution (25\%) stages of the system workflow. Collectively, our findings provide a critical roadmap for developers seeking to design the next generation of reliable and robust AI coding tools. 
\end{abstract}


\received{28 September 2023}
\received[revised]{5 March 2024}
\received[accepted]{16 April 2024}

\maketitle

\section{Introduction}
\label{sec:intro}

The advent of powerful Large Language Models (LLMs) has fundamentally altered the software development landscape~\cite{wang2025agents,jin2024llms,abdollahi2025surveying,wang2024software}. Moving beyond simple code completion, tools such as Claude Code~\cite{Claude}, Codex~\cite{Codex}, and Gemini CLI~\cite{Gemini} provide interactive, conversational interfaces capable of interpreting natural language instructions, performing task planning, invoking tools, maintaining session memory, modifying existing codebases, and executing system commands. These tools promise to minimize boilerplate code, expedite debugging workflows, and democratize programming by lowering technical barriers. However, transforming these systems from prototype demonstrations into production-ready, reliable developer tools requires substantial engineering effort. 

Existing research has largely focused on evaluating the outputs of AI coding tools, such as the correctness, efficiency, and security of the code generated by these tools~\cite{chatlatanagulchai2025use,santos2025decoding,finnie2022robots,brennan2022exploring,koksaldi2025accuracy}. In contrast, the engineering of the tool layers that wrap, integrate, and operationalize these models has received far less attention. Building an effective AI coding tool is fundamentally a systems engineering challenge: it requires embedding unstable and non-deterministic AI components into otherwise deterministic software systems with stringent requirements for reliability, security, and user experience~\cite{sculley2015hidden}. Despite the widespread enthusiasm for these tools, we still have limited empirical understanding of how well they are engineered in practice, how reliable they are, and what classes of bugs and failures they tend to exhibit.  

To bridge this gap, in this work, we study three widely used AI coding tools, namely Claude Code~\cite{Claude}, Codex~\cite{Codex}, and Gemini CLI~\cite{Gemini}, as representative examples for investigating engineering defects that arise during the development and deployment of AI-assisted programming tools. These tools maintain publicly accessible issue trackers that contain rich, real-world bug reports submitted by users while operating the tools in local development environments. The reported issues span a broad spectrum of engineering defects and reveal key challenges in building reliable AI coding systems, including functional errors, environment and dependency misconfigurations, and unexpected behaviors exhibited by the underlying LLMs. Collectively, these issues provide valuable empirical evidence for understanding the unique engineering challenges introduced by integrating LLMs into practical software development workflows. 
Specifically, we conduct the first systematic analysis of over 3.8K issues collected from the three AI coding tools that were closed by December 12, 2025, to answer the following research questions:

\textbf{RQ1 (Bug Type): What types of bugs occur in these tools?} This RQ categorizes the bugs found in the studied tools by their nature to characterize the overall defect landscape of AI coding tools.  

\textbf{RQ2 (Root Cause): What are the root causes of these bugs?} This RQ explores the underlying causes of bugs, aiming to provide a deeper understanding of why they arise in AI coding tools. 


\textbf{RQ3 (Bug Symptom): What are the symptoms of these bugs?} This RQ examines how bugs manifest to users at runtime, aiming to understand how underlying engineering defects translate into user-visible failures and affect developer workflows.  

\textbf{RQ4 (Bug Location): Which architectural layers do bugs occur in?} This RQ examines how reported bugs are distributed across the six-layer architecture of a typical AI coding tool (Figure~\ref{arcchart}) to identify how different bug categories manifest throughout the system.  


We found that most bugs in AI coding tools are related to core functionalities (i.e., functional bugs, 67\%), while a subset of bugs reflects user concerns regarding ease of use (usability and UI bugs, 17.9\%) and model compatibility (7.6\%). The primary root causes of these issues are engineering-related, such as configuration and integration errors. These bugs manifest through multiple symptom types, with tool/API errors and command/terminal failures dominating the distribution. Consequently, they primarily impact the external tool orchestration and command execution layers of the system, highlighting critical points of vulnerability in AI coding workflows. 
By revealing the engineering pitfalls in existing AI coding tools, this study provides actionable insights for designing more reliable, usable, and robust AI coding tools. 
This work makes the following contributions:

\begin{itemize}
    \item We conduct the first systematic empirical analysis of over 3.8K real-world issues collected from three widely used AI coding tools, i.e., Claude Code, Codex, and Gemini CLI, providing a comprehensive characterization of bugs encountered in practical AI-powered programming workflows.
    
    \item We propose a structured taxonomy that categorizes bugs according to their type, locations, observable symptoms, and root causes, offering a systematic framework for understanding defects in AI coding tools. 

    \item We quantify the prevalence and impact of different bug categories across tools, highlighting systematic reliability weaknesses and tool-specific failure profiles. 
   
    \item Based on our findings, we derive actionable implications for designing and testing AI coding tools.
\end{itemize}

\noindent \textbf{Data Availability:} Our data are publicly available at 
\url{https://zenodo.org/records/18342487}.

The remainder of this paper is organized as follows.  Section~\ref{sec:background} provides the background for our study. Section~\ref{sec:setup} describes the design and setup of our empirical study. Section~\ref{sec:result} presents the study results, while Section~\ref{sec:dis} discusses the key lessons learned. Section~\ref{sec:threats} outlines the potential threats to validity, and Section~\ref{sec:conclusion} concludes the paper.

\section{Background and Related Work}
\label{sec:background}

\subsection{AI Coding Tools}
\label{sec:2.1}

\begin{figure}[t!]  
    \centering
    \includegraphics[width=0.45\textwidth]{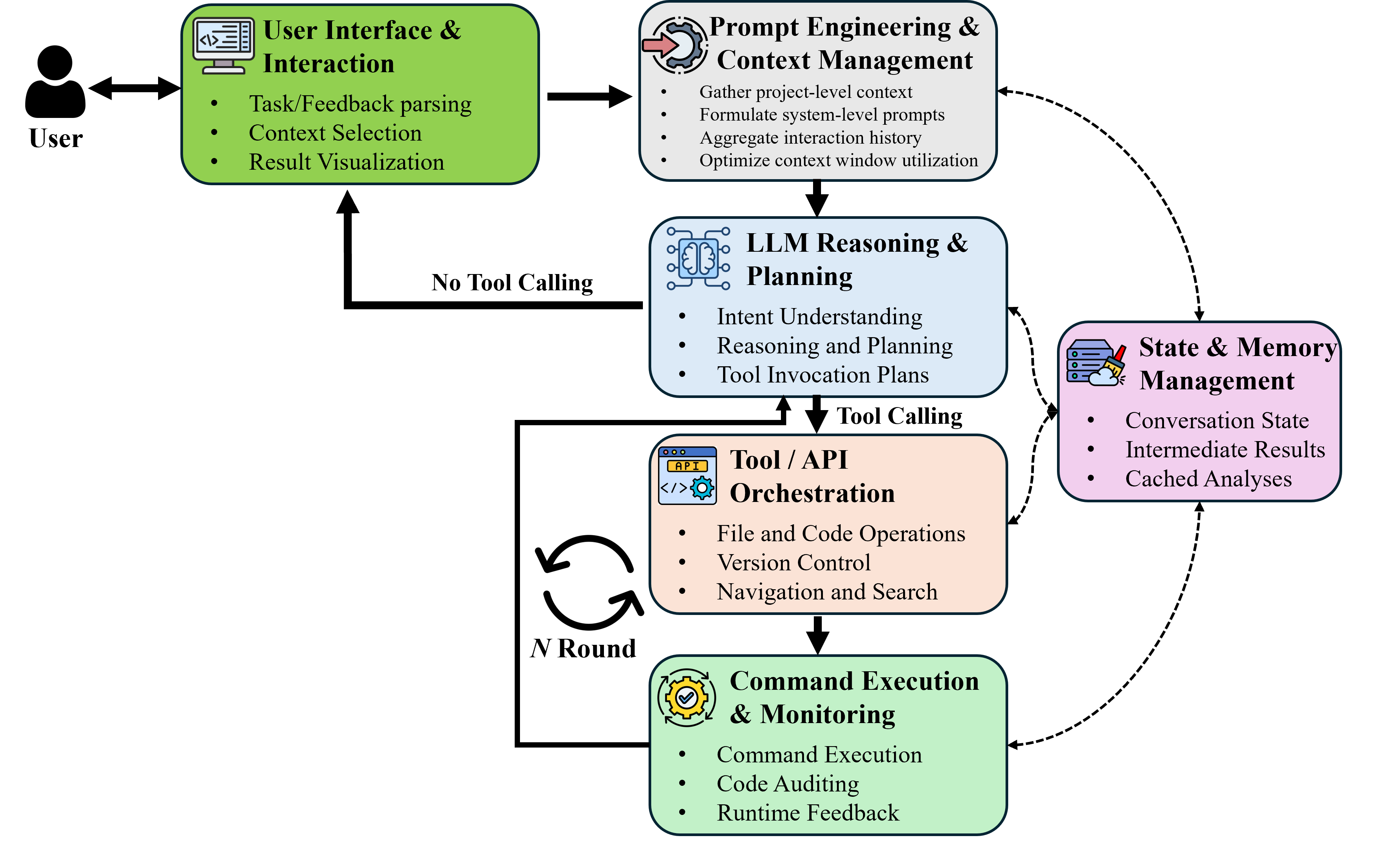} 
    \vspace{-0.1in}
    \caption{An Overview of AI Coding Tool}
    \label{arcchart}
\end{figure}
AI coding tools, such as Claude Code~\cite{Claude}, Codex~\cite{Codex}, and Gemini CLI~\cite{Gemini}, represent a new paradigm in AI-assisted software development. These systems function as autonomous coding agents that integrate tightly with developers' workflows through command-line interfaces and IDE extensions. Despite differences in implementation, these systems share a common design paradigm: they leverage LLMs as their core reasoning engines to interpret natural-language commands, decompose high-level goals into executable steps, and coordinate tool usage to carry out multi-step programming tasks. By combining the reasoning capabilities of LLMs with program analysis, tool orchestration, and execution feedback, AI coding tools can perform tasks with limited human intervention while remaining responsive to developer input and oversight. We reviewed the online development documentation of the three tools and distilled their architectures into a common pattern, as illustrated in Figure~\ref{arcchart}. 
The system's architecture 
comprises six main distinct components that collectively enable sophisticated autonomous behavior, which are as follows. 

\subsubsection{User Interface \& Interaction}
This component serves as the primary interface between developers and the Claude Code system. It captures user commands, retrieves the current editor state (e.g., open files, cursor position, selected text), and displays results back to users through various UI components. 

\subsubsection{Prompt Engineering \& Context Management}
This component governs how the AI coding tool translates user intent into effective model instructions. It dynamically constructs prompts by combining user requests, relevant code snippets, project context, system directives, and tool metadata. Because LLMs are highly sensitive to context framing, this layer also handles context compression, token budgeting, retrieval of relevant files, and conflict resolution between multiple sources of intent (e.g., user, system, agent). 

\subsubsection{LLM Reasoning \& Planning} This component forms the foundation of AI coding tools and governs their interaction with the underlying LLM backends. It encompasses API invocation, streaming token handling, output parsing, rate limiting, and error handling, including retries and fallback strategies across different models or decoding configurations. Beyond basic communication, this component is responsible for orchestrating reasoning and planning behaviors, such as task decomposition, step ordering, and context management. 

\subsubsection{Tool / API Orchestration}
AI coding assistants act as agents that rely on structured tool calls such as file editing, shell execution, code search, running tests, and interacting with version control to translate natural language directives into real system actions. The tool-invocation component defines the schema for these tools, interprets LLM-issued commands, validates parameters, and routes requests to the appropriate backend capabilities. It maintains a registry of all available tools and intelligently dispatches them based on the LLM’s decisions. 

\subsubsection{Command Execution \& Monitoring}
This component provides the runtime environment in which code is compiled, executed, or tested. It includes interpreters, language runtimes, containerized sandboxes, build systems, dependency managers, and process-control mechanisms. The LLM relies on the correctness of this component to run commands, analyze outputs, and refine code. 

\subsubsection{State \& Memory Management}

This component maintains the state of ongoing conversations and tasks, implementing both short-term working memory and long-term persistence. It tracks what has been done, what needs to be done, which files have been modified, and what errors have occurred. It often also includes a prompt caching mechanism to reduce token usage by caching frequently accessed context like project structure and documentation.

\subsection{IDE Design for AI-Powered Programming}
\label{sec:2.2}

Research on Integrated Development Environment (IDE) design has long examined how tooling, interaction paradigms, and information presentation affect developer productivity and experience. Early work by Wasserman et al.~\cite{wasserman1987graphical} introduced a graphical, extensible IDE architecture emphasizing modularity and customization, laying the groundwork for viewing IDEs as integrated platforms rather than simple editors. Subsequent studies explored IDE evaluation and adoption, highlighting context-dependent usability, developer expertise variability, and incremental evolution driven by workflow needs~\cite{kline2005evaluation, hou2009empirical}.

Beyond feature-centric design, novel interaction paradigms such as Code Bubbles~\cite{bragdon2010code} demonstrated that spatially organized, task-focused interfaces can reduce cognitive load and improve navigation. With the rise of AI-assisted programming, IDE research has increasingly focused on human–AI interaction. Studies have identified challenges in trust, controllability, error handling, and understanding in natural-language-driven code generation~\cite{xu2022ide}, while recent surveys synthesized design patterns and evaluation gaps for LLM-integrated IDEs~\cite{sergeyuk2024ide, sergeyuk2025human, treude2025developers}. Parallel work links IDE design, developer experience, and productivity, emphasizing transparency, feedback, and cognitive support as key factors~\cite{razzaq2024systematic, kumar2025intuition}, and exploring AI-driven automation in system architecture~\cite{anasuri2023ai}.

\section{Empirical Study Setup}
\label{sec:setup}

\subsection{Studied AI Coding Tools}
\label{sec:3.1}

In this study, we examine Claude Code~\cite{Claude}, Codex~\cite{Codex}, and Gemini CLI~\cite{Gemini} to investigate the engineering bugs that arise in modern agentic programming tools. These tools implement the common agentic workflow and are sufficiently complex to surface real-world engineering failures. 

\subsection{Data Collection}
\label{sec:3.2}

We collect the bugs reported by end users for each of the three tools from their publicly accessible GitHub repositories, ensuring transparency and reproducibility. The detailed steps are as follows.

\noindent \textbf{Step 1: Automatic Issue Retrieval.} For each repository, we used the GitHub REST API to retrieve the complete history of closed issues labeled as bugs. GitHub issues may represent different types of requests, such as new features, tests, debugging tasks, documentation, or bug reports; we therefore filtered issues by type to focus specifically on bugs. We considered only closed issues to ensure that each report included sufficient discussion, resolution context, or confirmation of a defect. 
For each issue, we collected and stored metadata including the title, description, labels, comments, timestamps, and resolution status for subsequent analysis. 

\noindent \textbf{Step 2: Manual Verification and Deduplication.}  
Since labeling practices vary across repositories, we used label-based filtering only as an initial heuristic rather than a definitive criterion. We then manually inspected each remaining issue to ensure data quality. During this process, we removed: (1) issues marked as duplicates of earlier reports that described the same underlying problem; (2) deprecated issues associated with obsolete software versions, experimental features that were later abandoned, or components that had been officially discontinued; (3) test or mock issues intentionally created by developers for purposes such as debugging, regression testing, continuous integration validation, or demonstrating issue-tracking workflows; and (4) issues labeled as bugs that did not correspond to actual defects, including reports caused by user misuse or configuration errors. 
The remaining issues constitute our final bug dataset.

\begin{table}[t!]
    \caption{The number of issues from each study tool}
    \vspace{-0.1in}
    \centering
    \label{tab:data}
\begin{tabular}{l|c|c|c}
\hline
Tool &\#Total Issues& \#Closed Bugs & \#Studied Bugs \\ \hline \hline
Claude Code  &  14,804 &   3,536         &      2,343    \\ \hline
  Gemini CLI   & 7,988   &     618      &      329        \\ \hline
   Codex   & 3,883    &   1,897       &       1,192        \\ \hline
\end{tabular}
\end{table}

The first two authors were involved in the manual inspection process and conducted their assessments independently. In cases of disagreement, two additional authors participated in discussions to reach a consensus and finalize the decision. To quantify the reliability of the manual labeling process, we measured inter-rater agreement using Cohen’s kappa, which yielded a value of 0.82, indicating a high level of agreement. The number of bugs retained at each step of the filtering process is shown in Table~\ref{tab:data}. 

\subsection{Data Labeling}
\label{sec:3.3}
To construct a comprehensive taxonomy of issues in AI coding assistants, three of the authors work together employ an open-card sorting approach. This methodology allows categories to emerge inductively from the data rather than being imposed a priori.

We begin by randomly sampling issues from each of the three tools to construct initial categories. Based on standard sample size estimation~\cite{jones2003introduction}, at a 95\% confidence level with a 5\% margin of error and an assumed population proportion of 20\%, the representative sample sizes are 223 instances for Claude Code, 141 for Gemini, and 204 for Codex. 
We therefore select 200 issues per tool to (i) ensure an equal number of samples across tools for fair comparison and (ii) provide sufficient coverage for empirical characterization within each tool. This yields an initial corpus of 600 issues for subsequent analysis. 

For each sampled issue, we manually examine the issue title, issue description, and associated comments to understand the reported problem, its manifestation, and the surrounding discussion. This manual analysis allows us to capture both surface-level symptoms and underlying engineering causes reflected in real-world usage. The three authors collaboratively developed the initial categories through thorough discussion and iterative refinement.  

Based on this analysis, we iteratively constructed an initial taxonomy along four orthogonal dimensions: (1) Bug Type, describing the category of the defect; (2) Root Cause, capturing the underlying engineering or architectural reason for the issue; (3) Symptom, characterizing how the issue manifests from the user or system perspective; and (4) Bug Location, indicating the stage of the AI coding assistant (see Section~\ref{sec:2.1}) in which the issue occurs (e.g., user interaction, prompt engineering, LLM reasoning, tool orchestration, execution, or memory management). 
We then abstract recurring patterns across issues to derive high-level categories and, where appropriate, finer-grained subcategories within each dimension. This iterative process continues until the taxonomy stabilizes and no new major categories emerge within the sampled dataset.

After establishing the initial taxonomy, we apply it to label the remaining issues outside the initial sample. During this labeling phase, if the existing taxonomy cannot adequately capture an issue, we refine or extend the taxonomy by introducing new categories or subcategories. 

\section{Result Analysis}
\label{sec:result}

In this section, we present and discuss our analysis results to address the four research questions we asked in Section~\ref{sec:intro}.

\subsection{RQ1: Bug Types}
\label{4.1}

\begin{table}[t!]
\centering
\caption{Bug type distribution by model}
\vspace{-0.1in}
\label{tab:bugtype-by-model}
\resizebox{0.48\textwidth}{!}{  
\begin{tabular}{lrrrrr}
\hline
Bug Type & Claude Code & Gemini & Codex CLI & Total & Rate (\%) \\
\hline
Functional Bugs     & 1567   & 225 & 795 & 2587 & 67.0 \\
Usability \& UI Bugs &  422 &  48 & 220 &  690 & 17.9 \\
Compatibility Bugs  &  147 &  30 & 117 &  294 &  7.6 \\
Performance Bugs    &  170 &  19 &  38 &  227 &  5.9 \\
Security Bugs       &   37 &   7 &  22 &   66 &  1.7 \\
\hline
\textbf{Total} & \textbf{2343} & \textbf{329} & \textbf{1192} & \textbf{3864} & \textbf{100.0} \\
\hline
\end{tabular}
}
\end{table}

We categorize these bugs into five types, i.e., \textit{Functionality Bug}, \textit{Usability \& UI Bug}, \textit{Compatibility Bug}, \textit{Performance Bug}, and \textit{Security Bugs}.   
Table~\ref{tab:bugtype-by-model} shows their distributions across the three tools. Detailed definitions of each type, along with examples, are provided below. 


\noindent \textbf{Functionality Bug:} is the bug that prevents one or more core capabilities from working as intended, leading to incorrect or incomplete task execution~\cite{tan2014bug}. This category includes cases where a tool invocation, command, or model-assisted operation produces wrong outputs, fails to apply changes, or does not follow the expected workflow logic. 
Common examples include edit operations that fail to update files as requested, as well as file search or navigation features that return incorrect results or fail to execute. 
For example, in Codex issue \href{https://github.com/openai/codex/issues/593}{\#593}, a user reported that Codex fails to execute the \textit{apply\_patch} command and instead merely displays the patch text. This is a functional bug since a core feature (i.e., automatic patch application) is not working as expected. The LLMs' inherent inaccuracies (generating wrong output or misunderstanding instructions) could not essentially be a bug, as it only shows the instability of the model's inner inference (prediction) logic, and it is rarely reported by users. However, if the AI coding tool consistently generates wrong outputs or misreads the instruction under some prompts with the same meaning, it could be a functional problem.


\noindent \textbf{Usability \& UI Bug:} 
A Usability \& UI bug impairs a user’s ability to interact effectively with AI coding tools by introducing friction, ambiguity, or misleading feedback. This category includes interface-level issues such as flickering, misaligned layouts, missing elements, or incorrect UI states that prevent users from accurately interpreting or manipulating code suggestions. It also encompasses higher-level usability problems, including confusing workflows, unclear instructions, inconsistent prompts, or misleading documentation, which increase cognitive load and hinder task completion. Such bugs can lead to errors, reduce productivity, and influence navigation, ultimately reducing efficiency and overall user experience when interacting with these AI-assisted coding systems. 
For example, in Gemini issue \href{https://github.com/google-gemini/gemini-cli/issues/7016}{\#7016}, users requested improved handling of content overflow in the terminal interface, replacing verbose truncated output (\textit{... X lines truncated}) with robust visual scrollbars. The existing overflow behavior makes it difficult to read long outputs or navigate content efficiently, indicating a UI concern that affects readability and interaction.

 
\noindent  \textbf{Compatibility Bug:} A compatibility bug arises from platform- or environment-specific constraints, where system behavior differs across operating systems, dependency versions, or terminal environments. Such bugs are characterized by features that function correctly in one environment but fail or behave differently in another. For example, a feature may fail on Windows due to version incompatibilities or exhibit incorrect input behavior under specific configurations. 
In Claude Code issue \href{https://github.com/anthropics/claude-code/issues/300}{\#300}, a user experienced a ``\textit{node: bad option}'' error when trying to run Claude Code on Ubuntu. This issue is caused by an unsupported Node.js version, which leads to a compatibility bug.

\noindent  \textbf{Performance Bug:} A performance bug degrades efficiency in terms of latency, throughput, or resource usage, even if the tool still produces correct results. This category includes high computational costs, redundant operations, unintended recursion, and high memory consumption that significantly slow or crash workflows. 
In Claude Code issue \href{https://github.com/anthropics/claude-code/issues/336}{\#336}, a user reported a crash due to an OutOfMemory error when running Claude Code in a specific codebase. This is a classic performance issue that crashes the system. 

\noindent  \textbf{Security Vulnerability:} A security vulnerability introduces vulnerabilities or unsafe behaviors that may enable unauthorized access, data exposure, or unintended destructive actions. This includes improper path or workspace validation (e.g., bypass via filesystem constructs), insecure handling of authentication state, and tool behaviors that execute sensitive operations without adequate safeguards. For example, a workspace restriction can be bypassed to access unintended files, or a destructive repository operation is performed without proper permission confirmation. Moreover, in Claude Code issue \href{https://github.com/anthropics/claude-code/issues/263}{\#263}, users reported that running or updating the claude CLI as the superuser (\textit{root}) poses a security risk. Specifically, the tool’s installation and authentication instructions encourage operations requiring elevated privileges (e.g., \textit{chmod -R} or \textit{chown -R}), which can destabilize systems or expose critical files when executed with root permissions. 

According to 
Table~\ref{tab:bugtype-by-model},  
bugs related to functionality account for the largest proportion (67\%), indicating that many reported issues involve feature failures or incorrect/unexpected outputs during usage. Usability \& UI and compatibility bugs represent the second and third largest categories (17.9\% and 7.6\%, respectively), suggesting that users often encounter interaction problems that require extra steps, cause confusion in workflows, or raise concerns about stability and consistent behavior. Together, these three categories make up approximately 85\% of all bug reports.  Besides those frequent bugs, security bugs appear the least (only 1.7\%), but they might lead to serious problems such as private data leakage.

\mybox{\textbf{Answer to RQ1:}
Functional bugs are the most frequent and directly impede users from using the tools. While usability \& UI, compatibility,  and performance bugs occur less often, they still impact the user experience. Security vulnerabilities are the least common, but their potential consequences are severe.} 


\subsection{RQ2: Root Cause}
\label{4.2}

\newcolumntype{P}[1]{>{\raggedright\arraybackslash}p{#1}}

\begin{table*}[t!]
\centering
\scriptsize
\setlength{\tabcolsep}{3pt}
\renewcommand{\arraystretch}{1.05}
\vspace{-0.15in}
\caption{Root cause taxonomy for observed failures (categories, sub-categories, and descriptions).}
\label{tab:root_cause_taxonomy}
\begin{tabular}{|P{4.8cm}|P{8.6cm}|P{1.2cm}|P{1.2cm}|}
\hline
\textbf{Category \& Sub-categories} & \textbf{Description} & \textbf{No.} & \textbf{Rate} \\
\hline

\textbf{API \& Integration Errors}\newline
\textbullet\ API Structure / Validation\newline
\textbullet\ Key \& Authentication Handling\newline
\textbullet\ Network / Proxy / VPN Issues\newline
\textbullet\ Third-Party Integration\newline
\textbullet\ Malformed Input Handling
&
Failures occur at system integration boundaries when interacting with external APIs or services.
\newline APIs reject requests due to schema mismatches or invalid parameters.
\newline Authentication tokens or API keys are missing, expired, or mis-scoped.
\newline Requests fail due to network routing, proxy, or VPN constraints.
\newline Third-party services behave inconsistently or are incompatible.
\newline Malformed or unexpected inputs cause request rejection or undefined behavior.
&
\textbf{826}\newline
325\newline
242\newline
103\newline
82\newline
74
&
\textbf{21.4\%}\newline
8.4\%\newline
6.3\%\newline
2.7\%\newline
2.1\%\newline
1.9\%
\\
\hline

\textbf{Configuration \& Setup}\newline
\textbullet\ Misconfigured Tools / Files\newline
\textbullet\ Path \& Directory Problems\newline
\textbullet\ Environment Variable Issues\newline
\textbullet\ Installation \& Update Problems\newline
\textbullet\ Documentation Errors
&
Failures caused by incorrect, incomplete, or inconsistent system configuration.
\newline Tools or configuration files are misconfigured or incompatible.
\newline Incorrect directory layouts or path resolution lead to runtime errors.
\newline Required environment variables are missing or mis-set.
\newline Installation, upgrade, or rollback processes fail.
\newline Documentation inaccuracies cause incorrect user setup.
&
\textbf{613}\newline
186\newline
132\newline
127\newline
104\newline
64
&
\textbf{15.9\%}\newline
4.8\%\newline
3.4\%\newline
3.3\%\newline
2.7\%\newline
1.6\%
\\
\hline

\textbf{User Interaction \& UI}\newline
\textbullet\ Input / Output Handling\newline
\textbullet\ UI Rendering / Display\newline
\textbullet\ Keyboard Shortcuts\newline
\textbullet\ Keybinding Conflicts\newline
\textbullet\ Clipboard \& Data Leak
&
Failures arising from incorrect assumptions about user interaction or interface behavior.
\newline User inputs or outputs are mishandled or misinterpreted.
\newline UI components render incorrectly or inconsistently.
\newline Keyboard shortcuts do not function as expected.
\newline Conflicting keybindings disrupt workflows.
\newline Clipboard usage leads to unintended data exposure.
&
\textbf{430}\newline
190\newline
145\newline
63\newline
25\newline
7
&
\textbf{11.1\%}\newline
4.9\%\newline
3.8\%\newline
1.6\%\newline
0.7\%\newline
0.2\%
\\
\hline

\textbf{Environment \& Platform Compatibility}\newline
\textbullet\ OS / Platform Restrictions\newline
\textbullet\ Shell Incompatibility\newline
\textbullet\ Terminal Emulator Issues\newline
\textbullet\ Node Version Issues\newline
\textbullet\ File System Differences
&
Failures due to mismatches between system assumptions and execution environments.
\newline Platform-specific constraints prevent correct execution.
\newline Shell differences cause command failures.
\newline Terminal emulators behave inconsistently.
\newline Incompatible runtime versions cause errors.
\newline File system semantics differ across environments.
&
\textbf{404}\newline
154\newline
118\newline
64\newline
34\newline
34
&
\textbf{10.5\%}\newline
4.0\%\newline
3.1\%\newline
1.7\%\newline
0.9\%\newline
0.9\%
\\
\hline

\textbf{AI Logic \& Behavior}\newline
\textbullet\ Instruction Noncompliance\newline
\textbullet\ Incorrect Output\newline
\textbullet\ Model / Subscription Recognition\newline
\textbullet\ Feature Hallucination\newline
\textbullet\ Content Filtering
&
Failures intrinsic to model reasoning or behavioral constraints.
\newline The model fails to follow explicit instructions.
\newline Generated outputs are logically incorrect or misleading.
\newline Model or subscription capabilities are misidentified.
\newline Hallucinated features or APIs are referenced.
\newline Content filtering interferes with task completion.
&
\textbf{385}\newline
145\newline
122\newline
80\newline
22\newline
16
&
\textbf{10.0\%}\newline
3.8\%\newline
3.1\%\newline
2.1\%\newline
0.6\%\newline
0.4\%
\\
\hline

\textbf{Performance \& Resource Limits}\newline
\textbullet\ Memory / Context Exceeded\newline
\textbullet\ Timeouts\newline
\textbullet\ Infinite Loops / Recursion
&
Failures caused by exceeding practical runtime or resource constraints.
\newline Context windows or memory limits are exceeded.
\newline Tasks terminate due to execution timeouts.
\newline Unintended infinite loops or recursion prevent completion.
&
\textbf{313}\newline
163\newline
121\newline
29
&
\textbf{8.0\%}\newline
4.1\%\newline
3.1\%\newline
0.8\%
\\
\hline

\textbf{Command Execution \& Parsing}\newline
\textbullet\ Process Control\newline
\textbullet\ Command Parsing\newline
\textbullet\ Argument Formatting\newline
\textbullet\ Argument Conflict / Logic\newline
\textbullet\ Bash / CLI Syntax Handling
&
Failures during command construction, parsing, or execution.
\newline Process lifecycle or control is mishandled.
\newline Commands are parsed incorrectly.
\newline Arguments are malformed or incompatible.
\newline Logical conflicts occur between command arguments.
\newline Shell syntax is not handled correctly.
&
\textbf{262}\newline
95\newline
63\newline
45\newline
36\newline
23
&
\textbf{6.8\%}\newline
2.5\%\newline
1.6\%\newline
1.2\%\newline
0.9\%\newline
0.6\%
\\
\hline

\textbf{State \& Context Management}\newline
\textbullet\ Session Persistence Issues\newline
\textbullet\ State Reset / Corruption\newline
\textbullet\ Context Compaction Errors\newline
\textbullet\ Instruction / Memory Handling\newline
\textbullet\ Conversation History Loss
&
Failures related to incorrect handling of session state or conversational context.
\newline Sessions fail to persist across interactions.
\newline Internal state resets or becomes corrupted.
\newline Context compaction removes critical information.
\newline Instruction or memory handling is flawed.
\newline Conversation history is lost unexpectedly.
&
\textbf{253}\newline
99\newline
72\newline
38\newline
26\newline
18
&
\textbf{6.6\%}\newline
2.6\%\newline
1.9\%\newline
1.0\%\newline
0.7\%\newline
0.5\%
\\
\hline

\textbf{Code \& File Handling}\newline
\textbullet\ File State Management\newline
\textbullet\ Line Endings \& Encoding\newline
\textbullet\ Large File Issues\newline
\textbullet\ Whitespace / Tab Handling
&
Failures caused by incorrect processing or manipulation of code and files.
\newline File state changes are tracked incorrectly.
\newline Encoding or line-ending mismatches occur.
\newline Large files exceed handling capabilities.
\newline Whitespace or tab inconsistencies cause errors.
&
\textbf{218}\newline
137\newline
51\newline
17\newline
13
&
\textbf{5.6\%}\newline
3.5\%\newline
1.3\%\newline
0.4\%\newline
0.3\%
\\
\hline

\textbf{Permissions \& Security}\newline
\textbullet\ Insufficient Permissions\newline
\textbullet\ Security Restrictions\newline
\textbullet\ Sensitive Data Handling\newline
\textbullet\ Superuser Misuse\newline
\textbullet\ Service Role Permissions
&
Failures related to improper enforcement or use of security mechanisms.
\newline Operations fail due to insufficient privileges.
\newline Platform security restrictions block actions.
\newline Sensitive data is mishandled.
\newline Elevated privileges are misused.
\newline Service role permissions are incorrectly scoped.
&
\textbf{160}\newline
85\newline
48\newline
12\newline
8\newline
7
&
\textbf{4.1\%}\newline
2.2\%\newline
1.2\%\newline
0.3\%\newline
0.2\%\newline
0.2\%
\\
\hline

\end{tabular}
\end{table*}

\begin{table}[t]
\centering
\vspace{-0.1in}
\caption{Root cause category distribution by model.}
\label{tab:rootcause-by-model}
\resizebox{0.48\textwidth}{!}{  
\begin{tabular}{lrrrrr}
\hline
Root Cause Category & Claude Code & Gemini & Codex CLI & Total & Rate (\%) \\
\hline
API \& Integration Errors      & 495 & 49 & 282 & 826 & 21.4 \\
Configuration \& Setup   & 437 & 41 & 135 & 613 & 15.9 \\
User Interaction \& UI       & 232 & 61 & 137 & 430 & 11.1 \\
Environment \& Platform Compatibility     & 231 & 43 & 130 & 404 & 10.5 \\
AI Logic \& Behavior & 159 & 43 & 183 & 385 & 10.0 \\
Performance \& Resource Limits    & 200 & 21 &  92 & 313 &  8.1 \\
Command Execution \& Parsing    & 180 & 18 &  64 & 262 &  6.8 \\
State \& Context Management    & 171 & 17 &  65 & 253 &  6.5 \\
Code \& File Handling    & 145 & 26 &  47 & 218 &  5.6 \\
Permissions \& Security &  93 & 10 &  57 & 160 &  4.1 \\
\hline
\textbf{Total} & \textbf{2343} & \textbf{329} & \textbf{1192} & \textbf{3864} & \textbf{100.0} \\
\hline
\end{tabular}
}
\end{table}

To systematically analyze the underlying reasons behind observed failures, we construct a root cause taxonomy for the issues studied. Each root cause was determined via evidence-based manual inspection: for each issue, we reviewed the title, description, and full discussion thread and extracted concrete evidence such as error logs/stack traces, command outputs and exit codes, and reproduction steps. 
As illustrated in Table.~\ref{tab:root_cause_taxonomy}, the taxonomy organizes failures into high-level root cause categories and finer-grained subcategories.
The taxonomy of root causes is categorized as follows:

\noindent \textbf{API \& Integration Errors:} This root cause category covers failures that occur at integration points, where the system interacts with external APIs or third-party components. Subcategories include mismatches in API structure or validation logic, integration incompatibilities, network or proxy constraints, improper handling of authentication keys, and malformed or unexpected inputs exchanged between systems.  

\noindent \textbf{Configuration \& Setup:}
  Root causes in this category stem from incorrect, incomplete, or inconsistent configuration during system setup or maintenance. Subcategories include problems related to environment variables, directory structures and paths, installation or update processes, tool or configuration file settings, as well as documentation that leads users to configure the system incorrectly.

\noindent \textbf{User Interaction \& UI:}
Root causes in this category arise from incorrect assumptions about user interaction and interface behavior. Subcategories include improper handling of input and output, unsupported or conflicting keyboard shortcuts, incorrect UI rendering or display states, unsafe clipboard interactions, and keybinding conflicts that disrupt expected workflows. 

\noindent \textbf{Environment \& Platform Compatibility:}
  This category arises from mismatches between system and execution environment, including differences in shells, operating systems, runtime versions, terminal emulators, and file system semantics.

\noindent \textbf{AI Logic \& Behavior:} This category captures root causes intrinsic to AI-driven components and model behavior. Subcategories include systematically incorrect outputs, hallucinated assumptions, unintended effects of content filtering, failure to follow instructions, and incorrect assumptions about available models or subscription-level capabilities.

\noindent \textbf{Performance \& Resource Limits:}
Root causes in this category arise when a workflow unexpectedly exceeds runtime-, configuration-, or platform-imposed resource limits, preventing tasks from completing. Subcategories include exceeding execution time limits (timeouts), exhausting memory or context budgets, and pathological computation patterns (e.g., non-terminating loops or unbounded recursion) that lead to resource exhaustion and prevent tasks from completing.

\noindent \textbf{Incorrect Command Execution \& Parsing:}
This category encompasses issues stemming from errors in command construction, parsing, or execution. Subcategories include incorrect handling of Bash or CLI syntax, command parsing failures, improper argument formatting, flawed process control, and logical conflicts between command arguments. 

\noindent \textbf{Improper State \& Context Management:} Root causes in this category arise from improper handling of session state, conversational context, or persistent memory. Subcategories include loss of session persistence, missing or truncated conversation history, errors during context compaction, incorrect instruction or memory handling, and unintended state resets or corruption.

\noindent \textbf{Code \& File Handling:} Root causes in this category relate to how source code and files are processed, modified, or managed by the system. Subcategories include inconsistencies in line endings or character encoding, whitespace or tab handling errors, incorrect tracking of file state, and limitations or failures when processing large files.

\noindent \textbf{Permissions \& Security:} This category arises from improper enforcement or usage of permission and security mechanisms. Subcategories include insufficient or overly restrictive access rights, misuse of elevated privileges, platform-imposed security restrictions, unsafe handling of sensitive data, and incorrectly scoped service roles or permission models.
Table~\ref{tab:rootcause-by-model} presents the aggregated distribution of root causes of the studied bugs grouped by high-level root cause categories, with counts shown separately for each tool. 
The primary root causes of bugs are dominated by \textit{API \& Integration Errors} and \textit{Configuration \& Setup}, which account for 21.4\% and 15.9\% of all the bugs observed, respectively. These categories underscore persistent challenges at integration boundaries and in environment configuration across all three tools. Other notable causes include \textit{User Interaction \& UI} and \textit{Environment \& Platform Compatibility}, which account for 11.1\% and 10.5\% of all the bugs observed, respectively, reflecting issues stemming from assumptions about user workflows and execution contexts. 
Bugs attributed to \textit{AI Logic \& Behavior} account for only 10\% of all observed bugs; this category of root causes indicates intrinsic limitations of model reasoning and constraint adherence, though they are not the predominant type. 
Less frequent, but still significant, are \textit{Permissions \& Security} and \textit{Code \& File Handling}, which exhibit model-dependent variation.

\mybox{\textbf{Answer to RQ2:} 
Our root cause analysis shows that bugs primarily originate from systemic engineering issues. Such as configuration and setup errors, API and integration mismatches, and environment or platform incompatibilities, rather than from isolated coding mistakes or LLM reasoning errors.}

\subsection{RQ3: Bug Symptom}
\label{4.3}
\newcolumntype{P}[1]{>{\raggedright\arraybackslash}p{#1}}
\begin{table*}[t!]
\centering
\scriptsize
\setlength{\tabcolsep}{3pt}
\renewcommand{\arraystretch}{1.05}
\caption{Symptom categories and sub-categories}
\label{tab:symptom_taxonomy}
\begin{tabular}{|c|P{4.8cm}|P{8.6cm}|P{1.0cm}|P{1.0cm}|}
\hline
\textbf{ID} & \textbf{Category-Subcategory} & \textbf{Description} & \textbf{No.} & \textbf{Rate} \\
\hline
1 &
\textbf{API \& Server Communication}\newline
\textbullet\ API Errors \& Status Codes\newline
\textbullet\ Connection/Timeout Issues\newline
\textbullet\ Invalid/Blocked Requests\newline
\textbullet\ Rate Limits \& Quotas\newline
\textbullet\ API Schema \& Contract Errors
&
Remote requests fail or return unexpected server-side errors during interaction with backend services.\newline
The system returns explicit API failures with status codes.\newline
Requests stall or time out before the service responds.\newline
Requests are rejected due to validation errors, policy blocks, or forbidden endpoints.\newline
Operations are terminated after exceeding request quotas or usage limits.\newline
Request or response fields do not match the expected API schema.
&
\textbf{708}\newline
403\newline
124\newline
37\newline
90\newline
54
&
\textbf{18.3\%}\newline
10.4\%\newline
3.2\%\newline
0.9\%\newline
2.3\%\newline
1.4\%
\\
\hline
2 &
\textbf{CLI/Terminal Experience}\newline
\textbullet\ UI Rendering Problems\newline
\textbullet\ Input \& Keybinding Issues\newline
\textbullet\ Terminal Output Errors\newline
\textbullet\ Autocomplete \& Reference Issues\newline
\textbullet\ IME/Locale-Related Input Issues
&
Terminal-based interaction behaves unexpectedly, making input, output, or navigation unreliable.\newline
The terminal UI renders incorrectly, with layout glitches or flicker.\newline
Key presses and shortcuts are ignored or mapped to unintended actions.\newline
Terminal output is missing, truncated, or corrupted.\newline
Autocomplete suggestions or references point to incorrect or incomplete contents.\newline
Locale or IME composition breaks command entry or text input.
&
\textbf{542}\newline
197\newline
188\newline
66\newline
51\newline
40
&
\textbf{14.0\%}\newline
5.1\%\newline
4.9\%\newline
1.7\%\newline
1.3\%\newline
1.0\%
\\
\hline
3 &
\textbf{Command Execution}\newline
\textbullet\ Timeouts \& Hanging\newline
\textbullet\ Shell \& Terminal Bugs\newline
\textbullet\ Unsupported Commands\newline
\textbullet\ Interrupted Sessions\newline
\textbullet\ Autocomplete \& Keybinding Failures
&
Commands executed on behalf of users do not run to completion or behave incorrectly.\newline
Invoked commands hang or exceed expected execution time.\newline
Shell or terminal issues that cause command failures.\newline
Required commands are unavailable or not allowed in the current environment.\newline
Execution is terminated during running.\newline
Completion or shortcut-driven control fails within execution workflows.
&
\textbf{492}\newline
216\newline
106\newline
125\newline
37\newline
8
&
\textbf{12.7\%}\newline
5.6\%\newline
2.7\%\newline
3.2\%\newline
0.9\%\newline
0.2\%
\\
\hline
4 &
\textbf{Code \& File Operations}\newline
\textbullet\ File Editing \& Replacement\newline
\textbullet\ File/Folder Structure\newline
\textbullet\ Context Management\newline
\textbullet\ Search \& Indexing\newline
\textbullet\ Symlink \& Symbolic Link Handling
&
Editing project files produces missing, misplaced, or inconsistent changes.\newline
Edits are not applied correctly or modify the wrong files.\newline
Files or directories are created, moved, or referenced at incorrect paths.\newline
The system applies changes using the wrong file context or loses needed context.\newline
Search results are incomplete, irrelevant, or miss the intended files.\newline
Operations fail or misbehave when paths traverse symbolic links.
&
\textbf{281}\newline
121\newline
62\newline
52\newline
31\newline
14
&
\textbf{7.3\%}\newline
3.1\%\newline
1.6\%\newline
1.4\%\newline
0.8\%\newline
0.4\%
\\
\hline
5 &
\textbf{AI Model \& Output Control}\newline
\textbullet\ Harmful Content Generation\newline
\textbullet\ Overly Affirmative/Unintended Responses\newline
\textbullet\ Unauthorized Modifications\newline
\textbullet\ Misleading/Incorrect AI Output\newline
\textbullet\ Simulated/Non-Functional Execution
&
Model issues that confuse users or derail task completion.\newline
Generated content violates safety expectations or includes disallowed contents.\newline
The model produces inappropriate agreement or unintended behavior.\newline
Actions are taken beyond the user’s stated intent or scope.\newline
Advice or generated code is incorrect in ways that take users toward failure.\newline
The system says actions were executed but actually not.
&
\textbf{271}\newline
7\newline
17\newline
38\newline
164\newline
45
&
\textbf{7.0\%}\newline
0.2\%\newline
0.4\%\newline
1.0\%\newline
4.2\%\newline
1.2\%
\\
\hline
6 &
\textbf{Authentication \& Access}\newline
\textbullet\ OAuth \& Login Errors\newline
\textbullet\ API Key Issues\newline
\textbullet\ Permissions/Authorization
&
Login, credential, or permission flows fail, preventing access to required features or resources.\newline
OAuth flows fail with redirects, loops, or missing tokens.\newline
API keys are missing, invalid, or rejected by the system.\newline
Requests fail due to insufficient privileges or incorrect authorization scope.
&
\textbf{258}\newline
143\newline
59\newline
56
&
\textbf{6.7\%}\newline
3.7\%\newline
1.5\%\newline
1.5\%
\\
\hline
7 &
\textbf{User Feedback \& Usability}\newline
\textbullet\ Misleading Output\newline
\textbullet\ Documentation Gaps\newline
\textbullet\ Logging \& Notifications\newline
\textbullet\ Keyboard Shortcuts \& Keybindings Issues\newline
\textbullet\ Focus \& Multitasking Problems
&
System messages and guidance are unclear or misleading, increasing user effort.\newline
Status messages report is inconsistent with actual outcomes.\newline
Help text or documentation omits information needed to proceed correctly.\newline
Logs or notifications are absent or noisy.\newline
Global shortcuts behave inconsistently, increasing interaction cost.\newline
Focus management breaks when users switch tasks or contexts.
&
\textbf{243}\newline
151\newline
38\newline
27\newline
12\newline
15
&
\textbf{6.2\%}\newline
3.9\%\newline
1.0\%\newline
0.7\%\newline
0.3\%\newline
0.4\%
\\
\hline
8 &
\textbf{Session \& State Management}\newline
\textbullet\ Session Persistence\newline
\textbullet\ Unsaved Progress \& Task Resumption\newline
\textbullet\ Task \& Agent State Loss\newline
\textbullet\ Context Continuity \& Restoration\newline
\textbullet\ Session Timeouts \& Expiry
&
Sessions fail to persist, resume, or maintain consistent internal state across interactions.\newline
Sessions are not saved or restored.\newline
Sessions that in progress can not be resumed from a previous checkpoint.\newline
Agent state resets unexpectedly, losing progress or decisions.\newline
Prior context is not carried forward, causing inconsistent follow up behavior.\newline
Sessions expire or timeout prematurely.
&
\textbf{224}\newline
62\newline
14\newline
71\newline
65\newline
12
&
\textbf{5.8\%}\newline
1.6\%\newline
0.4\%\newline
1.9\%\newline
1.7\%\newline
0.3\%
\\
\hline
9 &
\textbf{Configuration \& State}\newline
\textbullet\ Settings Persistence\newline
\textbullet\ Env Variables \& Paths\newline
\textbullet\ Update/Restart Loops\newline
\textbullet\ State Persistence Issues\newline
\textbullet\ Configuration File Bloat
&
Configurations are applied inconsistently, causing repeated failures across runs.\newline
User settings revert or fail to take effect consistently.\newline
Environment variables or path resolution are missing or incorrect.\newline
Repeated updates or restarts prevent stable use of the system.\newline
Internal state is not persisted correctly between runs.\newline
Configuration artifacts grow or corrupt, degrading stability or usability.
&
\textbf{209}\newline
74\newline
77\newline
28\newline
11\newline
19
&
\textbf{5.4\%}\newline
1.9\%\newline
2.0\%\newline
0.7\%\newline
0.3\%\newline
0.5\%
\\
\hline
10 &
\textbf{Installation \& Setup}\newline
\textbullet\ Platform Compatibility\newline
\textbullet\ Permissions \& Privileges\newline
\textbullet\ Package Management\newline
\textbullet\ Shell/Environment Issues
&
Setup steps fail during environment preparation, blocking initial use or correct operation.\newline
Behavior differs across OS/architecture, causing setup or runtime failures.\newline
Operations fail due to missing privileges or restricted access.\newline
Package installation fails due to conflicts or version mismatches.\newline
Shell or environment configuration prevents correct execution.
&
\textbf{166}\newline
42\newline
34\newline
52\newline
38
&
\textbf{4.3\%}\newline
1.1\%\newline
0.9\%\newline
1.3\%\newline
1.0\%
\\
\hline
11 &
\textbf{Security \& Data Management}\newline
\textbullet\ Credential Leaks\newline
\textbullet\ Silent/Unauthorized Actions\newline
\textbullet\ File/Directory Access Control\newline
\textbullet\ Keychain \& Credential Store Issues
&
Security handling is flawed, risking exposure, unauthorized actions, or incorrect access control.\newline
Secrets appear in outputs or are stored in insecure locations.\newline
Actions occur without authorization.\newline
Restricted files or directories can be accessed when they should not be.\newline
Secure credential stores fail to read, write, or unlock secrets.
&
\textbf{128}\newline
8\newline
41\newline
33\newline
46
&
\textbf{3.3\%}\newline
0.2\%\newline
1.1\%\newline
0.9\%\newline
1.2\%
\\
\hline
12 &
\textbf{Integration \& Compatibility}\newline
\textbullet\ Third-party Tool Support\newline
\textbullet\ Editor Integration\newline
\textbullet\ Dev Container/WSL Issues\newline
\textbullet\ Proxy/VPN/Networking
&
Interactions with external tools break due to integration-specific constraints.\newline
External tools cannot be invoked or coordinated correctly.\newline
IDE/editor workflows break due to integration defects.\newline
Containerized or WSL environments trigger path, permission, or runtime mismatches.\newline
Network intermediaries disrupt connectivity or tool access.
&
\textbf{116}\newline
24\newline
39\newline
28\newline
25
&
\textbf{3.0\%}\newline
0.6\%\newline
1.0\%\newline
0.7\%\newline
0.7\%
\\
\hline
13 &
\textbf{Model Selection \& Availability}\newline
\textbullet\ Invalid Model Name Errors\newline
\textbullet\ Model Overload \& Availability\newline
\textbullet\ Custom/Default Model Resolution Errors
&
Model selection or access fails, preventing users from running with the model.\newline
Provided model identifiers are rejected or it is invalid.\newline
Requests fail because the selected model is overloaded or unavailable.\newline
The system resolves to the wrong model despite user configuration.
&
\textbf{97}\newline
39\newline
20\newline
38
&
\textbf{2.5\%}\newline
1.0\%\newline
0.5\%\newline
1.0\%
\\
\hline
14 &
\textbf{Hook \& Automation Issues}\newline
\textbullet\ Hook Configuration Errors\newline
\textbullet\ Automation Execution Failures\newline
\textbullet\ Session/Directory-Specific Execution Issues
&
Automation triggers run at the wrong time or fail to execute in the expected context.\newline
Hook definitions are malformed or misconfigured.\newline
Automations do not trigger or terminate unexpectedly.\newline
Automations behave differently depending on directory or session scope.
&
\textbf{53}\newline
20\newline
23\newline
10
&
\textbf{1.4\%}\newline
0.5\%\newline
0.6\%\newline
0.3\%
\\
\hline
15 &
\textbf{Cost \& Billing}\newline
\textbullet\ Unexpected Charges\newline
\textbullet\ Cost Reporting Errors
&
Billing or subscription-related issues that undermine trust in charging and accounting.\newline
Users observe charges that do not align with their expected usage or plan.\newline
Usage dashboards or cost summaries reports are incorrect.
&
\textbf{39}\newline
22\newline
17
&
\textbf{1.0\%}\newline
0.6\%\newline
0.4\%
\\
\hline
16 &
\textbf{Image \& Media Handling}\newline
\textbullet\ Image Upload/Processing Errors\newline
\textbullet\ Image Display Issues
&
Media upload, processing, or rendering fails, preventing the successful use of image-related features.\newline
Uploads fail or processing stalls.\newline
Images fail to render correctly or do not appear in the UI.
&
\textbf{37}\newline
33\newline
4
&
\textbf{0.9\%}\newline
0.8\%\newline
0.1\%
\\
\hline
\end{tabular}

\end{table*}

This question investigates how bugs manifest to users during runtime. By categorizing these observable symptoms, we aim to understand how underlying engineering defects translate into user-facing failures and affect developer workflows. 




We identified 16 categories of bug symptoms from the observed bugs, each comprising several subcategories. Details of these categories are provided in Table~\ref{tab:symptom_taxonomy}. Below, we summarize the symptom taxonomy and describe the typical user-visible behaviors associated with each type of issue. 

\noindent \textbf{API \& Server Communication.} Symptoms in this category are primarily manifested as failures in communicating with APIs, where requests are rejected, delayed, or return unexpected errors. Subcategories include: 1) \textit{API Errors \& Status Codes}: explicit API error messages with status codes (e.g., 4xx/5xx) indicating request failures, 2) \textit{Connection/Timeout Issues}: requests time out, stall, or the server appears non-responsive, 3) \textit{Invalid/Blocked Requests}: requests are denied due to some reasons, 4) \textit{Rate Limits \& Quotas}: users hit usage limits and receive warnings, 5) \textit{API Schema \& Contract Errors}: mismatches between expected and actual request or response formats of APIs. 
  
\noindent  \textbf{CLI/Terminal Experience.} This category describes user-visible issues in terminal-based interactions, where the interface behaves unexpectedly or input/output is not rendered reliably. Subcategories include: 1) \textit{UI Rendering Problems}: terminal UI is misaligned, flickers, or displays corrupted layout, 2) \textit{Input \& Keybinding Issues}: keystrokes (e.g., arrows, shortcuts) are ignored or trigger wrong behavior, 3) \textit{Terminal Output Errors}: output is truncated or missing, 4) \textit{Autocomplete \& Reference Issues}: suggestions are incorrect, incomplete, or reference wrong files/commands, 5) \textit{IME/Locale-Related Input Issues}: non-English input methods cause broken typing, composition, or command entry. 
  
\noindent \textbf{Command Execution.} Symptoms in this category occur when the system runs commands on the user’s behalf, and execution does not proceed as expected, often preventing task completion. Subcategories include: 1) \textit{Timeouts \& Hanging}: commands stall indefinitely or exceed expected runtime, 2) \textit{Shell \& Terminal Bugs}: shell-specific behaviors (path parsing, etc.) lead to failures, 3) \textit{Unsupported Commands}: attempted commands are unavailable or not permitted, 4) \textit{Interrupted Sessions}: execution is terminated mid-run, 5) \textit{Autocomplete \& Keybinding Failures}: command-line completion or shortcut-driven control fails during execution workflows. 
  
\noindent \textbf{Code \& File Operations.} This category captures symptoms where code editing, file manipulation, or repository operations behave incorrectly, leading to missing changes or inconsistent project state. It includes: 1) \textit{File Editing \& Replacement}: edits are not applied, partially applied, or overwrite the wrong content, 2) \textit{File/Folder Structure}: incorrect paths, missing directories, or unexpected file placements, 3) \textit{Context Management}: the system uses the wrong context or loses necessary context during editing, 4) \textit{Search \& Indexing}: file search is inaccurate, incomplete, or returns irrelevant results, 5) \textit{Symlink \& Symbolic Link Handling}: operations fail or behave unexpectedly when symlinks are present. 
  
\noindent \textbf{AI Model \& Output Control.} Symptoms here relate to the model’s behavioral output rather than infrastructure failures, including unsafe, misleading, or non-actionable responses. Subcategories include: 1) \textit{Harmful Content Generation}: outputs violate safety expectations or generate content that is not allowed, 2) \textit{Overly Affirmative/Unintended Responses}: the model agrees incorrectly or produces unintended responses, 3) \textit{Unauthorized Modifications}: the system performs changes beyond what the user requested, 4) \textit{Misleading/Incorrect AI Output}: generated guidance or code is incorrect in a way that misleads the user, 5) \textit{Simulated/Non-Functional Execution}: the model claims actions were performed but actually not, or produces hallucinated results. 
  
\noindent \textbf{Authentication \& Access.} This category reflects symptoms in identity, credential, and permission workflows that prevent users from accessing features or resources. Subcategories include: 1) \textit{OAuth \& Login Errors}: login redirects fail, tokens are not issued, or sign-in loops occur; 2) \textit{API Key Issues}: keys are missing, rejected, invalid, or not recognized; 3) \textit{Permissions/Authorization}: requests fail due to insufficient privileges or incorrect authorization scope. 
  
\noindent \textbf{User Feedback \& Usability.} Symptoms in this category reduce user confidence and efficiency by providing unclear, incomplete, or misleading guidance about system status and required actions. Subcategories are: 1) \textit{Misleading Output}: messages imply success/failure incorrectly or provide confusing messages, 2) \textit{Documentation Gaps}: missing, outdated, or ambiguous instructions that block progress, 3) \textit{Logging \& Notifications}: logs are absent, or do not show actionable error details, 4) \textit{Keyboard Shortcuts \& Keybindings Issues}: shortcuts are inconsistent or not functional, increasing interaction cost, 5) \textit{Focus \& Multitasking Problems}: context switching is fragile (e.g., losing focus, difficulty managing multiple tasks). 
  
\noindent \textbf{Session \& State Management.} This category describes failures to preserve or restore ongoing work, leading to lost progress or broken continuation. Subcategories include: 1) \textit{Session Persistence}: sessions are not saved or restored, 2) \textit{Unsaved Progress \& Task Resumption}: unfinished work cannot be resumed from a prior session, 3) \textit{Task \& Agent State Loss}: internal agent state resets unexpectedly, 4) \textit{Context Continuity \& Restoration}: prior context is not carried over, causing inconsistent follow-up behavior, 5) \textit{Session Timeouts \& Expiry}: sessions expire too early that interrupt workflows.

\begin{table}[t]
\centering
\caption{Symptom category distribution by model.}
\vspace{-0.1in}
\label{tab:symptom-category-by-model}
\resizebox{\linewidth}{!}{ 
\begin{tabular}{lrrrrr}
\hline
Symptom Category & Claude Code & Gemini & Codex CLI & Total & Rate (\%) \\
\hline
API \& Server Communication     & 423 & 40 & 245 & 708 & 18.3 \\
CLI/Terminal Experience         & 316 & 72 & 154 & 542 & 14.0 \\
Command Execution               & 288 & 40 & 164 & 492 & 12.7 \\
Code \& File Operations         & 171 & 36 &  74 & 281 &  7.3 \\
AI Model \& Output Control      & 112 & 27 & 132 & 271 &  7.0 \\
Authentication \& Access        & 129 & 24 & 105 & 258 &  6.7 \\
User Feedback \& Usability      & 164 & 22 &  57 & 243 &  6.3 \\
Session \& State Management     & 140 & 16 &  68 & 224 &  5.8 \\
Configuration \& State          & 145 & 15 &  49 & 209 &  5.4 \\
Installation \& Setup           & 122 & 10 &  34 & 166 &  4.3 \\
Security \& Data Management     & 107 &  5 &  16 & 128 &  3.3 \\
Integration \& Compatibility    &  68 & 10 &  38 & 116 &  3.0 \\
Model Selection \& Availability &  59 &  3 &  35 &  97 &  2.5 \\
Hook \& Automation Issues       &  48 &  5 &   0 &  53 &  1.4 \\
Cost \& Billing                 &  28 &  3 &   8 &  39 &  1.0 \\
Image \& Media Handling         &  23 &  1 &  13 &  37 &  1.0 \\
\hline
\textbf{Total} & \textbf{2343} & \textbf{329} & \textbf{1192} & \textbf{3864} & \textbf{100.0} \\
\hline
\end{tabular}
}
\end{table}

\noindent \textbf{Configuration \& State.} Symptoms here arise when configuration settings or runtime state are applied inconsistently, resulting in repeated failures or unstable behavior across runs. Include: 1) \textit{Settings Persistence}: user settings do not persist, revert, or apply inconsistently, 2) \textit{Env Variables \& Paths}: environment variables or path resolution are incorrect or missing, 3) \textit{Update/Restart Loops}: repeated update prompts or restart cycles, 4) \textit{State Persistence Issues}: internal state is not correctly stored or recovered, 5) \textit{Configuration File Bloat}: config artifacts grow unexpectedly that become corrupted or degrade usability. 
  
\noindent \textbf{Installation \& Setup.} This category captures symptoms during initial setup or environment preparation, where prerequisites fail and prevent the system from running correctly. Subcategories are: 1) \textit{Platform Compatibility}: installation fails or behaves differently across operating system or architecture, 2) \textit{Permissions \& Privileges}: insufficient permissions block installs, tool access, or required operations, 3) \textit{Package Management}: dependency installation fails, conflicts occur, or versions mismatch, 4) \textit{Shell/Environment Issues}: incorrect shell configuration, missing environment activation, or incompatible shell behavior.
  
\noindent \textbf{Security \& Data Management.} Symptoms in this category involve security risks or incorrect handling of sensitive resources, potentially exposing data or performing actions without adequate safeguards. Subcategories include: 1) \textit{Credential Leaks}: secrets appear in logs/output or are stored insecurely, 2) \textit{Silent/Unauthorized Actions}: actions are performed without proper authorization, 3) \textit{File/Directory Access Control}: improper access to restricted files or directories, 4) \textit{Keychain \& Credential Store Issues}: failures reading or writing credentials from secure stores. 
  
\noindent \textbf{Integration \& Compatibility.} This category shows symptoms that appear when the system interacts with external tools or runs under varied development environments, where integrations break or behave inconsistently. Subcategories include: 1) \textit{Third-party Tool Support}: failures invoking or coordinating external tools, 2) \textit{Editor Integration}: problems occur when integrating with IDE/editor workflows, 3) \textit{Dev Container/WSL Issues}: failures in containerized or WSL environments, 4) \textit{Proxy/VPN/Networking}: networking intermediaries disrupt connectivity or tool access. Common symptoms that we usually meet are related to errors in using models in IDEs such as VSCode and Cursor. 
  
\noindent \textbf{Model Selection \& Availability.} Symptoms here occur when users select models or when the platform resolves model choices incorrectly, leading to being unable to run to select models. This category include: 1) \textit{Invalid Model Name Errors}: model identifiers (i.e., name) are rejected or unrecognized, 2) \textit{Model Overload \& Availability}: capacity limits or outages prevent model access, 3) \textit{Custom/Default Model Resolution Errors}: the system chooses the wrong model (e.g., ignores user selection or chooses wrong defaults).
  
\noindent \textbf{Hook \& Automation Issues.} Symptoms related to automation logic (hooks, scripted triggers, scheduled actions) that fail to run reliably or run in the wrong context. Subcategories are: 1) \textit{Hook Configuration Errors}: hook definitions are invalid or misconfigured, 2) \textit{Automation Execution Failures}: automation does not trigger, crashes, or ends early, 3) \textit{Session/Directory-Specific Execution Issues}: automation behaves differently depending on working directory or session scope. 
  
\noindent \textbf{Cost \& Billing.} This category is reflected in user-observed billing issues. Such as users observing charges that do not match expectations, or usage dashboards and cost summaries are incorrect.
  
\noindent \textbf{Image \& Media Handling.} This category describes symptoms arising when users upload, process, or view media, where media operations fail or render incorrectly. Subcategories include \textit{Image Upload/Processing Errors}, where uploads fail, processing stalls, or formats are rejected unexpectedly, and \textit{Image Display Issues}, which images render incorrectly or do not appear.


We further show the distribution of symptom categories across the three tools in Table~\ref{tab:symptom-category-by-model}. \textit{API \& Server Communication} is the most frequent category (708, 18.3\%), followed by C\textit{LI/Terminal Experience} (542, 14.0\%) and \textit{Command Execution} (492, 12.7\%). Together, these three categories account for roughly 45\% of all observed symptoms, indicating that user-visible failures often emerge when the system interacts with outside tools, terminal-based interaction, or executes commands.


Beyond these leading categories, several mid-frequency symptom groups, such as \textit{Code \& File Operations} (281, 7.3\%), \textit{AI Model \& Output Control} (271, 7.0\%), and \textit{Authentication \& Access} (258, 6.7\%) remain non-trivial, reflecting that users also encounter failures in local project manipulation, model output, and access workflows. In contrast, categories such as \textit{Cost \& Billing} (39, around 1.0\%) and \textit{Image \& Media Handling} (37, less than 1.0\%) appear comparatively rare in the data, suggesting that user-visible issues are dominated by inner execution and interaction rather than peripheral platform features. 

\mybox{\textbf{Answer to RQ3:} Our symptom analysis indicates that failures are most visible during the system’s execution steps, particularly when it relies on outside tools and command-line operations. In comparison, symptoms that occur outside the core execution pipeline occur less frequently.}

\subsection{RQ4: Bug Locations}
\label{4.4}




\begin{table}[t]
\centering
\caption{Bug location distribution by model.}
\vspace{-0.1in}
\label{tab:archlayer-by-model}
\resizebox{\linewidth}{!}{
\begin{tabular}{lrrrrr}
\hline
Architectural Layer & Claude Code & Gemini & Codex CLI & Total & Rate (\%) \\
\hline
Tool/API Orchestration            & 949 & 98 & 408 & 1455 & 37.6 \\
Command Execution \& Monitoring   & 563 & 68 & 332 &  963 & 25.0 \\
User Interface \& Interaction     & 347 & 75 & 207 &  629 & 16.3 \\
State \& Memory Management        & 228 & 30 &  94 &  352 &  9.1 \\
LLM Reasoning Planning            & 113 & 22 & 111 &  246 &  6.4 \\
Prompt Engineering \& Context Management & 143 & 36 &  40 &  219 &  5.6 \\
\hline
\textbf{Total} & \textbf{2343} & \textbf{329} & \textbf{1192} & \textbf{3864} & \textbf{100.0} \\
\hline
\end{tabular}
}
\end{table}

\begin{figure}[t] 
    \centering
    \includegraphics[width=0.5\textwidth]{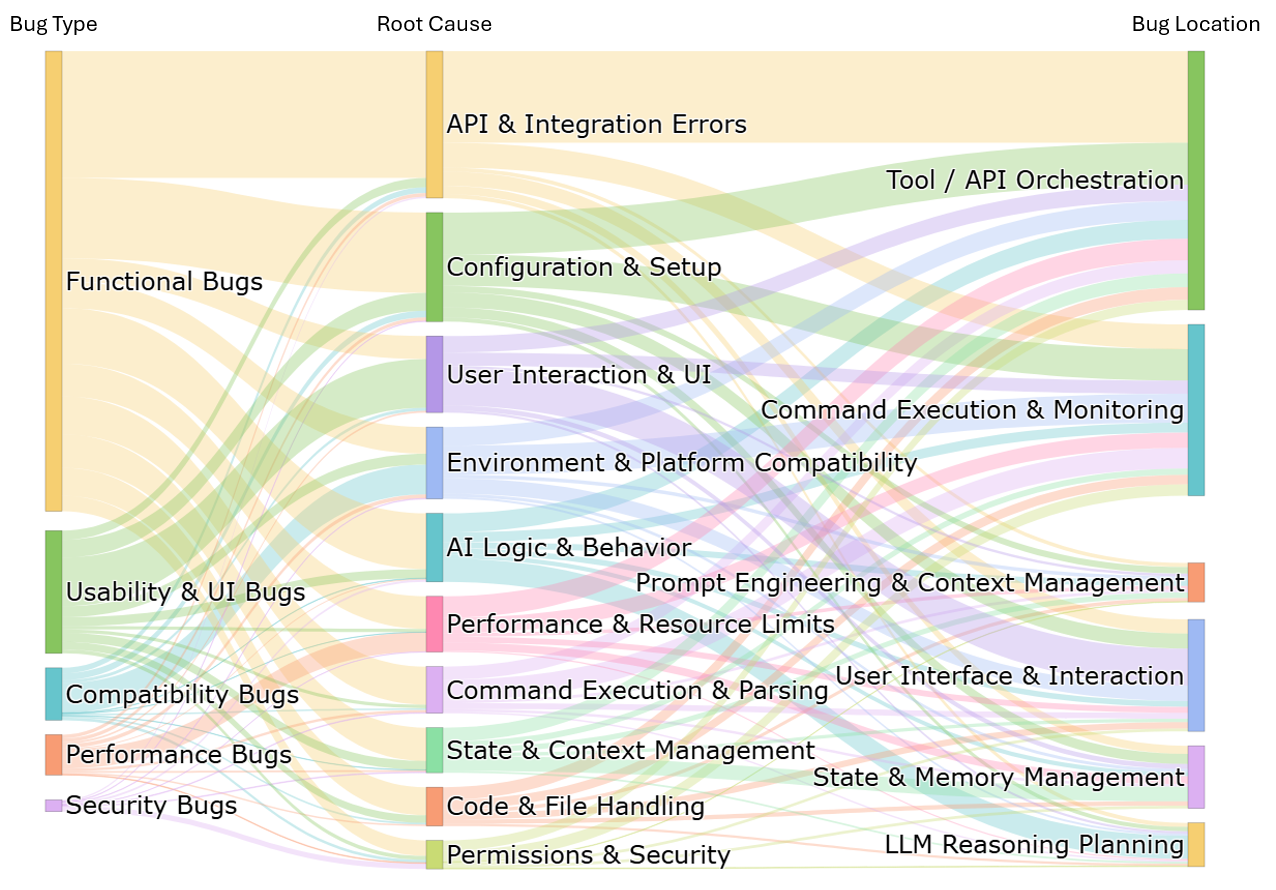}
    \caption{Mapping of bug types, root causes, and bug locations}
    \label{fig:mapping_of_bug_features}
\end{figure}
This research question aims to identify where failures manifest within the architectural stack of a typical AI coding tool (as shown in Section~\ref{sec:2.1}), thereby uncovering structural weaknesses, cross-layer dependencies, and components that are particularly prone to failure. Due to limited resources, we did not reproduce the studied bugs; instead, we inferred the bug location of each bug through careful analysis of bug reports, issue discussions, and code snippets when available. 
Consequently, we do not analyze how these failures propagate across layers.

Table~\ref{tab:archlayer-by-model} shows the bug distribution across different architectural layers of AI Coding tools.  
Our analysis reveals that bugs in LLMs are not uniformly distributed across architectural layers, but instead concentrate heavily in a small number of core layers. 
For example, \textit{Tool / API Orchestration} is the most affected layer, accounting for 37.6\% of studied bugs. 
This indicates that interactions between the AI system and external tools constitute a primary source of system fragility. 
A second major concentration of bugs occurs in \textit{Command Execution \& Monitoring}, reflecting the complexity of constructing, executing, and supervising system-level commands, which accounts for 25\% of all bugs.   
In contrast, \textit{User Interface \& Interaction} exhibits a moderate but consistent number of bugs, which accounts for 16.3\% of all bugs.    
These failures are typically localized and relate to mismatches between system assumptions and user behavior, such as unsupported inputs or incorrect rendering states, and rarely propagate deeply into other layers.

Figure~\ref{fig:mapping_of_bug_features} presents the mapping between bug types, their root causes, and the affected execution layers. Functional bugs dominate the distribution and exhibit a wide range of root causes; in particular, most bugs stemming from API misuse and configuration errors manifest as functional issues. In contrast, other bug types tend to have root causes closely aligned with their respective domains. For example, usability and UI bugs are primarily caused by user interaction or interface design issues and ultimately affect the UI layer. Moreover, bugs arising from diverse root causes frequently propagate to the execution stage of Tool/API orchestration, command execution, and UI. This observation reinforces our findings that core components of AI coding tools, such as rapid API invocation, command execution, and user interaction, remain key weaknesses, as these elements form the central logic of the tools and mediate the interaction between the system and its users.

\mybox{\textbf{Answer to RQ4:} 
Bugs in AI coding tools occur across multiple architectural layers. Issues in the \textit{Tool/API Orchestration} and \textit{Command Execution \& Monitoring} components are particularly prevalent, forming the dominant share of the observed failures.
}

\section{Lesson Learned}
\label{sec:dis} 
Our study reveals several interesting findings that can serve
as practical guidelines for both industry and academic communities to improve AI coding tool development in the future. 

\textbf{System-Level Failures Dominate}: 
Our results suggest that many failures in AI coding tools arise from weak coordination among the language model, supporting infrastructure, and execution environment, rather than model reasoning alone. As shown in Section~\ref{4.2}, \textit{API \& Integration Errors} (21.4\%) and \textit{Configuration \& Setup} issues (15.9\%) account for a large portion of failures, typically occurring at system boundaries such as API mismatches, authentication, network conditions, environment variables, path resolution, and version compatibility. Failures classified as \textit{AI Logic \& Behaviour}, which reflect intrinsic model issues like instruction noncompliance or hallucinations, represent a smaller share (10.0\%), likely due in part to reporting bias: developers tend to report failures that block workflows, while suboptimal model outputs that do not halt execution are often tolerated or corrected. Consequently, system-level failures that impact usability and workflow continuity dominate issue reports.  

\textbf{Explicitly Define and Maintain Interactions}: 
The interactions among LLMs, tools, and execution environments should be explicitly specified, validated, and continuously maintained. From a software engineering perspective, building AI coding tools introduces a substantially expanded set of artifacts, such as tool schemas, prompt templates, environment assumptions, and execution logs, all of which require systematic management and evolution. In addition, we also observe that many API and configuration failures stem from fragmentation issues, such as unexpected API formats, version mismatches, or incompatible runtime environments, which place significant stress on testing and validation. Proactively checking these conditions before execution can prevent cascading failures, avoid incorrect follow-up actions, and reduce user confusion. For example, AI coding tools could dynamically inspect users’ environments and automatically adapt or select compatible tools and configurations accordingly. 

\textbf{Provide Clear and Reliable Execution Feedback}: 
Execution feedback should be clear and reliable for both users and models. Many failures attributed to the tool were, in fact, caused by hidden errors or missing feedback from earlier command executions, which emphasizes the importance of proper logging and traceability. Providing clear and structured execution feedback helps both users and models understand what occurred, preventing incorrect attribution of failures to the model and supporting more effective debugging and workflow recovery. 


\textbf{Indicate Uncertain or Unreliable System States}:  Tools should clearly indicate when execution feedback or system state is unreliable. Without such signals, models may act on incorrect information and produce wrong outputs. Simple indicators of invalid or uncertain state can help models pause, retry, or ask for clarification, improving robustness and user trust. 
\section{Threats to Validity}
\label{sec:threats}
Our study is subject to several threats to validity. First, external validity may be limited by the scope of the studied tools. We analyze bugs from only three representative AI coding tools. Although these tools are widely used and cover different design choices, they may not fully represent the diversity of AI coding tools in terms of architectures, functionalities, and usage contexts. Consequently, our findings may not generalize to all AI-assisted coding systems. Second, construct validity is affected by the manual labeling process. The categorization of bug types, root causes, symptoms, and architectural locations was performed manually based on bug reports and related artifacts. While we followed a consistent labeling protocol, manual analysis is inherently subjective and may introduce bias or inconsistencies.

\section{Conclusion}
\label{sec:conclusion}

In this paper, we presented the first large-scale empirical study of real-world bugs in widely used AI coding tools, namely Claude Code, Codex CLI, and Gemini CLI. Through an analysis of over 3.8K user-reported bugs, we characterized the defect landscape across bug types, root causes, bug symptoms, and bug locations. Our results show that most reliability issues arise from integration, configuration, and usability challenges, rather than solely from limitations of LLM reasoning. These findings demonstrate that building reliable AI coding tools is fundamentally a systems engineering problem and highlight the need for robust integration strategies, defensive LLM interaction, and improved user-facing error handling.



\bibliographystyle{ACM-Reference-Format}
\bibliography{sample-acmsmall-conf}


\end{document}